# Simple Formulas for Generating Chern-Simons Basic Invariant Polynomials by Repeated Exterior Differentiation

C. C. Briggs
*Center for Academic Computing, Penn State University, University Park, PA 16802*

**Abstract.** Simple formulas are given for generating Chern-Simons basic invariant polynomials by repeated exterior differentiation for *n*-dimensional differentiable manifolds having a general linear connection.

PACS numbers: 02.40.-k, 04.20.Fy

This paper presents simple formulas for generating Chern-Simons basic invariant polynomials by repeated exterior differentiation for *n*-dimensional differentiable manifolds having a general linear connection.

The Chern-Simons basic invariant polynomials of such a manifold $M$ are the $(2p-1)$-forms $Q_p(\omega \wedge \Omega^{p-1})$ together with the $2p$-forms $Q_p(\Omega^p)$ defined, for $p > 0$, by[1]

$$\begin{aligned} Q_p(\omega \wedge \Omega^{p-1}) &= \omega_{i_p}{}^{i_1} \wedge \Omega_{i_1}{}^{i_2} \wedge \Omega_{i_2}{}^{i_3} \wedge \ldots \wedge \Omega_{i_{p-3}}{}^{i_{p-2}} \wedge \Omega_{i_{p-2}}{}^{i_{p-1}} \wedge \Omega_{i_{p-1}}{}^{i_p} \\ &= \mathrm{tr}(\omega \wedge \Omega^{p-1}) \end{aligned} \tag{1}$$

and

$$\begin{aligned} Q_p(\Omega^p) &= \Omega_{i_p}{}^{i_1} \wedge \Omega_{i_1}{}^{i_2} \wedge \Omega_{i_2}{}^{i_3} \wedge \ldots \wedge \Omega_{i_{p-3}}{}^{i_{p-2}} \wedge \Omega_{i_{p-2}}{}^{i_{p-1}} \wedge \Omega_{i_{p-1}}{}^{i_p} \\ &= \mathrm{tr}(\Omega^p), \end{aligned} \tag{2}$$

respectively, where $\omega_a{}^b$ and $\Omega_a{}^b$ are the connection 1-forms and curvature 2-forms, respectively, of $M$.

The $1^{st}$ ordinary exterior differentials of the basis tangent vectors $\mathsf{e}_a$ of $M$ are given by[2-4]

$$\mathrm{d}\, \mathsf{e}_a = \mathsf{e}_b\, \omega_a{}^b, \tag{3}$$

the contractions of which with the basis 1-forms $\omega^b$ of $M$ are given by

$$\langle \omega^b, \mathrm{d}\, \mathsf{e}_a \rangle = \omega_a{}^b \tag{4}$$

and in view of which the $1^{st}$ absolute exterior differentials of $\mathsf{e}_a$ are given by

$$\begin{aligned} \mathrm{D}\, \mathsf{e}_a &= \mathrm{d}\, \mathsf{e}_a - \mathsf{e}_b\, \omega_a{}^b \\ &= 0, \end{aligned} \tag{5}$$

where the contractions of $\omega^b$ with $\mathsf{e}_a$ are given by

$$\langle \omega^b, \mathsf{e}_a \rangle = \delta^b_a, \tag{6}$$

where $\delta^b_a$ is the Kronecker delta.

The $2^{nd}$ ordinary exterior differentials of $\mathsf{e}_a$ are given by[5-12]

$$\begin{aligned} \mathrm{d}^2\, \mathsf{e}_a &= \mathrm{d}\, \mathrm{d}\, \mathsf{e}_a \\ &= \mathrm{d}\, \mathsf{e}_b\, \omega_a{}^b \\ &= (\mathrm{d}\, \mathsf{e}_b) \wedge \omega_a{}^b + \mathsf{e}_b\, \mathrm{d}\, \omega_a{}^b \\ &= \mathsf{e}_c\, \omega_b{}^c \wedge \omega_a{}^b + \mathsf{e}_b\, \mathrm{d}\, \omega_a{}^b \\ &= \mathsf{e}_b\, (\mathrm{d}\, \omega_a{}^b + \omega_c{}^b \wedge \omega_a{}^c) \\ &= \mathsf{e}_b\, \Omega_a{}^b, \end{aligned} \tag{7}$$

the contractions of which with $\omega^b$ are given by

$$\langle \omega^b, \mathrm{d}^2\, \mathsf{e}_a \rangle = \Omega_a{}^b \tag{8}$$

and where

$$\begin{aligned} \Omega_a{}^b &= \mathrm{d}\, \omega_a{}^b + \omega_c{}^b \wedge \omega_a{}^c \\ &= \mathrm{D}\, \omega_a{}^b + \omega_a{}^c \wedge \omega_c{}^b \\ &= \mathrm{d}\, \omega_a{}^b - \omega_a{}^c \wedge \omega_c{}^b \\ &= \mathrm{D}\, \omega_a{}^b - \omega_c{}^b \wedge \omega_a{}^c. \end{aligned} \tag{9}$$

The $3^{rd}$ ordinary exterior differentials of $\mathsf{e}_a$ are given by

$$\begin{aligned} \mathrm{d}^3\, \mathsf{e}_a &= \mathrm{d}\, \mathrm{d}^2\, \mathsf{e}_a \\ &= \mathrm{d}\, \mathsf{e}_b\, \Omega_a{}^b \\ &= (\mathrm{d}\, \mathsf{e}_b) \wedge \Omega_a{}^b + \mathsf{e}_b\, \mathrm{d}\, \Omega_a{}^b \\ &= (\mathsf{e}_c\, \omega_b{}^c) \wedge \Omega_a{}^b + \mathsf{e}_b\, \mathrm{d}\, \Omega_a{}^b \\ &= \mathsf{e}_c\, \Omega_b{}^c \wedge \omega_a{}^b \end{aligned} \tag{10}$$

using Bianchi's identity for $\Omega_a{}^b$,[13] i.e.,

$$\begin{aligned} \mathrm{D}\, \Omega_a{}^b &= \mathrm{d}\, \Omega_a{}^b - \omega_a{}^c \wedge \Omega_c{}^b + \omega_c{}^b \wedge \Omega_a{}^c \\ &= 0, \end{aligned} \tag{11}$$

as well as by

$$\begin{aligned} \mathrm{d}^3\, \mathsf{e}_a &= \mathrm{d}^2\, \mathrm{d}\, \mathsf{e}_a \\ &= \mathrm{d}^2\, \mathsf{e}_b\, \omega_a{}^b \\ &= (\mathrm{d}^2\, \mathsf{e}_b) \wedge \omega_a{}^b + \mathsf{e}_b\, \mathrm{d}^2\, \omega_a{}^b \\ &= (\mathsf{e}_c\, \Omega_b{}^c) \wedge \omega_a{}^b + 0 \\ &= \mathsf{e}_c\, \Omega_b{}^c \wedge \omega_a{}^b \end{aligned} \tag{12}$$

1 Chern, S. S., and J. Simons, "Characteristic forms and geometric invariants," *Annals of Math.*, **99** (1974) 48.
2 Misner, C. W., K. S. Thorne, and J. A. Wheeler, *Gravitation*, W. H. Freeman & Co., San Francisco, CA (1973), p. 350.
3 Cartan, É. J., *Leçons sur la géométrie des espaces de Riemann*, Gauthier-Villars et C^ie^, Paris, France (1928), p. 180; *Leçons sur la géométrie des espaces de Riemann, 2nd édition, rev. et augm.*, Gauthier-Villars, Paris, France (1946), p. 180.
4 Chern, S. S., "A simple intrinsic proof of the Gauss-Bonnet formula for closed Riemannian manifolds," *Ann. of Math.*, **45** (1944) 747; "On the curvatura integra in a Riemannian manifold," *ibid.*, **46** (1945) 674; "On the Minimal Immersions of the Two-sphere in a Space of Constant Curvature," in Gunning, R. C. (ed.), *Problems in Analysis: A Symposium in Honor of Salomon Bochner*, Princeton University Press, Princeton, NJ (1970), p. 27.
5 Bott, R., and S. S. Chern, "Some Formulas Related to Complex Transgression," in Haefliger, A., and R. Narasimhan (eds.), *Essays on Topology and Related Topics; Memoires dédiés à Georges de Rham*, Springer-Verlag, New York, NY (1970), p. 48.
6 Cartan, É. J., "Sur les variétés à connexion affine et la théorie de la relativité généralisée (Première Partie)," *Annales scientifiques de l'École Normale supérieure*, **40** (1923) 325; (1928), *op. cit.*, p. 182; (1946), *op. cit.*, p. 181.
7 Chern, S. S., "Differential Geometry of Fiber Bundles," in *Proceedings of the International Congress of Mathematicians*, **2** (1950) 397; "Topics in Differential Geometry," [mimeographed notes], The Institute for Advanced Study, Princeton, NJ (1951), Chap. 3, Sec. 3; (1970), *op. cit.*
8 Cohen, J. M., *Exterior Differential Forms and Orthonormal Cartan Frames*, [unpublished monograph], (1975), p. 33.
9 Flanders, H., "Development of an Extended Exterior Differential Calculus," *Trans. Amer. Math. Soc.*, **75** (1953) 311; "Methods in Affine Connection Theory," *Pac. J. Math.*, **5** (1955) 391.
10 Misner, C. W., *et al.*, *op. cit.*, pp. 350-351 and 359.
11 Petrie, T., and J. Randall, *Connections, Definite Forms, and Four-manifolds*, Clarendon Press, Oxford, England (1990), pp. 11-12.
12 Trautmann, A., *Differential Geometry for Physicists—Stony Brook Lectures*, Bibliopolis, edizioni di filosofia e scienze, Napoli, Italy (1984), pp. 91-92.
13 Cartan, É. J., (1928), *op. cit.*, p. 214; (1946), *op. cit.*, p. 210.

using Poincaré's theorem for scalar-valued exterior differential forms,[14-15] i.e.,

$$\mathrm{d}^2\, \alpha = 0, \tag{13}$$

where $\alpha$ is an arbitrary scalar-valued exterior differential form.

The $4^{th}$ ordinary exterior differentials of $\mathrm{e}_a$ are given by

$$\begin{aligned} \mathrm{d}^4\, \mathrm{e}_a &= \mathrm{d}^2\, \mathrm{d}^2\, \mathrm{e}_a \\ &= \mathrm{d}^2\, \mathrm{e}_b\, \Omega_a{}^b \\ &= (\mathrm{d}^2\, \mathrm{e}_b) \wedge \Omega_a{}^b + \mathrm{e}_b\, \mathrm{d}^2\, \Omega_a{}^b \\ &= (\mathrm{e}_c\, \Omega_b{}^c) \wedge \Omega_a{}^b + 0 \\ &= \mathrm{e}_c\, \Omega_b{}^c \wedge \Omega_a{}^b. \end{aligned} \tag{14}$$

In general, the $p^{th}$ ordinary exterior differentials of $\mathrm{e}_a$ for $p > 0$ are given by

$$\begin{aligned} \mathrm{d}^p\, \mathrm{e}_a &= \begin{cases} \mathrm{e}_{i_1}\, \Omega_{i_2}{}^{i_1} \wedge \Omega_{i_3}{}^{i_2} \wedge \Omega_{i_4}{}^{i_3} \wedge \ldots \wedge \Omega_{i_{(p-1)/2}}{}^{i_{(p-3)/2}} \wedge \Omega_{i_{(p+1)/2}}{}^{i_{(p-1)/2}} \wedge \omega_a{}^{i_{(p+1)/2}}, & \text{if } p \text{ is odd} \\ \mathrm{e}_{i_1}\, \Omega_{i_2}{}^{i_1} \wedge \Omega_{i_3}{}^{i_2} \wedge \Omega_{i_4}{}^{i_3} \wedge \ldots \wedge \Omega_{i_{(p-2)/2}}{}^{i_{(p-4)/2}} \wedge \Omega_{i_{p/2}}{}^{i_{(p-2)/2}} \wedge \Omega_a{}^{i_{p/2}}, & \text{if } p \text{ is even} \end{cases} \\ &= \begin{cases} \mathrm{e}_{i_1}\, \omega_a{}^{i_{(p+1)/2}} \wedge \Omega_{i_{(p+1)/2}}{}^{i_{(p-1)/2}} \wedge \Omega_{i_{(p-1)/2}}{}^{i_{(p-3)/2}} \wedge \ldots \wedge \Omega_{i_4}{}^{i_3} \wedge \Omega_{i_3}{}^{i_2} \wedge \Omega_{i_2}{}^{i_1}, & \text{if } p \text{ is odd} \\ \mathrm{e}_{i_1}\, \Omega_a{}^{i_{p/2}} \wedge \Omega_{i_{p/2}}{}^{i_{(p-2)/2}} \wedge \Omega_{i_{(p-2)/2}}{}^{i_{(p-4)/2}} \wedge \ldots \wedge \Omega_{i_4}{}^{i_3} \wedge \Omega_{i_3}{}^{i_2} \wedge \Omega_{i_2}{}^{i_1}, & \text{if } p \text{ is even} \end{cases} \\ &= \begin{cases} \mathrm{e}_{i_{(p+1)/2}}\, \omega_a{}^{i_1} \wedge \Omega_{i_1}{}^{i_2} \wedge \Omega_{i_2}{}^{i_3} \wedge \ldots \wedge \Omega_{i_{(p-5)/2}}{}^{i_{(p-3)/2}} \wedge \Omega_{i_{(p-3)/2}}{}^{i_{(p-1)/2}} \wedge \Omega_{i_{(p-1)/2}}{}^{i_{(p+1)/2}}, & \text{if } p \text{ is odd} \\ \mathrm{e}_{i_{p/2}}\, \Omega_a{}^{i_1} \wedge \Omega_{i_1}{}^{i_2} \wedge \Omega_{i_2}{}^{i_3} \wedge \ldots \wedge \Omega_{i_{(p-6)/2}}{}^{i_{(p-4)/2}} \wedge \Omega_{i_{(p-4)/2}}{}^{i_{(p-2)/2}} \wedge \Omega_{i_{(p-2)/2}}{}^{i_{p/2}}, & \text{if } p \text{ is even} \end{cases}, \end{aligned} \tag{15}$$

the contractions of which with $\omega^a$ are given by

$$\begin{aligned} \langle \omega^a, \mathrm{d}^p\, \mathrm{e}_a \rangle &= \begin{cases} \delta_{i_{(p+1)/2}}{}^a\, \omega_a{}^{i_1} \wedge \Omega_{i_1}{}^{i_2} \wedge \Omega_{i_2}{}^{i_3} \wedge \ldots \wedge \Omega_{i_{(p-5)/2}}{}^{i_{(p-3)/2}} \wedge \Omega_{i_{(p-3)/2}}{}^{i_{(p-1)/2}} \wedge \Omega_{i_{(p-1)/2}}{}^{i_{(p+1)/2}}, & \text{if } p \text{ is odd} \\ \delta_{i_{p/2}}{}^a\, \Omega_a{}^{i_1} \wedge \Omega_{i_1}{}^{i_2} \wedge \Omega_{i_2}{}^{i_3} \wedge \ldots \wedge \Omega_{i_{(p-6)/2}}{}^{i_{(p-4)/2}} \wedge \Omega_{i_{(p-4)/2}}{}^{i_{(p-2)/2}} \wedge \Omega_{i_{(p-2)/2}}{}^{i_{p/2}}, & \text{if } p \text{ is even} \end{cases} \\ &= \begin{cases} \omega_{i_{(p+1)/2}}{}^{i_1} \wedge \Omega_{i_1}{}^{i_2} \wedge \Omega_{i_2}{}^{i_3} \wedge \ldots \wedge \Omega_{i_{(p-5)/2}}{}^{i_{(p-3)/2}} \wedge \Omega_{i_{(p-3)/2}}{}^{i_{(p-1)/2}} \wedge \Omega_{i_{(p-1)/2}}{}^{i_{(p+1)/2}}, & \text{if } p \text{ is odd} \\ \Omega_{i_{p/2}}{}^{i_1} \wedge \Omega_{i_1}{}^{i_2} \wedge \Omega_{i_2}{}^{i_3} \wedge \ldots \wedge \Omega_{i_{(p-6)/2}}{}^{i_{(p-4)/2}} \wedge \Omega_{i_{(p-4)/2}}{}^{i_{(p-2)/2}} \wedge \Omega_{i_{(p-2)/2}}{}^{i_{p/2}}, & \text{if } p \text{ is even} \end{cases} \\ &= \begin{cases} Q_{(p+1)/2}(\omega \wedge \Omega^{(p-1)/2}), & \text{if } p \text{ is odd} \\ Q_{p/2}(\Omega^{p/2}), & \text{if } p \text{ is even} \end{cases}. \end{aligned} \tag{16}$$

Equation (16) then yields the formulas in question, viz.,

$$Q_p(\omega \wedge \Omega^{p-1}) = \langle \omega^a, \mathrm{d}^{2p-1}\, \mathrm{e}_a \rangle \tag{17}$$

and

$$Q_p(\Omega^p) = \langle \omega^a, \mathrm{d}^{2p}\, \mathrm{e}_a \rangle. \tag{18}$$

For an aside, note that using Eq. (18) and taking the $0^{th}$ ordinary exterior differentials of $\mathrm{e}_a$ to be given by

$$\mathrm{d}^0\, \mathrm{e}_a = \mathrm{e}_a, \tag{19}$$

the polynomial $Q_p(\Omega^p)$ for $p = 0$ is given by

$$\begin{aligned} Q_0(\Omega^0) &= \langle \omega^a, \mathrm{d}^0\, \mathrm{e}_a \rangle \\ &= \langle \omega^a, \mathrm{e}_a \rangle \\ &= \delta^a_a \\ &= n, \end{aligned} \tag{20}$$

where $n$, having appeared above, is the number of dimensions of $M$.

14 Cartan, É. J., (1928), *op. cit.*, pp. 209 and 215; *Le systèmes différentiels extérieurs et leurs applications géometriques*, Actualités scientifiques et industrielles, 994, Exposés de géométrie, vol. 12, Librairie scientifiques Hermann & C$^{ie}$, Éditeurs, Paris, France (1945), p. 37; (1946), *op. cit.*, pp. 208 and 211.

15 Chern, S. S., "The Geometry of Isotropic Surfaces," *Annals of Math.*, **43** (1942) 545; "On the Euclidean Connections in a Finsler Space," *Proc. Nat. Acad. Sci.*, **29** (1943) 33; "Local Equivalence and Euclidean Connections in Finsler Spaces," *Science Reports Nat. Tsing Hua Univ.*, **5** (1948) 95.